\documentclass[journal]{IEEEtran}

\usepackage[]{graphicx}
\usepackage{caption}
\usepackage{subfig} %should come after caption package if given, can't be used with caption2, defines subfloat
\usepackage{amsmath}
\usepackage{amssymb}
\usepackage{epsfig}
\usepackage{cite}
\usepackage{color}
\usepackage{balance}
\usepackage{amsmath}
\usepackage{amsfonts} % for the \checkmark command 
\usepackage{graphicx}
\usepackage{pdfpages}
\usepackage[T1]{fontenc} % For less than and greater than sign
\usepackage{array}
\usepackage{url}

\usepackage{color}
\usepackage{wrapfig}
\usepackage{lipsum}
\usepackage[hidelinks]{hyperref}
\usepackage{multirow}
\usepackage{tabularx}
\usepackage{tikz}
\usepackage{booktabs}
\usepackage{textgreek}

\usepackage{algpseudocode} % For algorithm 
\usepackage{algorithm,algpseudocode}

\begin{document}

\title{Evaluating the Reliability of Air Temperature\\ from ERA5 Reanalysis Data}

\author{
Barry~McNicholl,
Yee~Hui~Lee,~\IEEEmembership{Senior~Member,~IEEE,}
Abraham G.~Campbell,~\IEEEmembership{Member,~IEEE}
and~Soumyabrata~Dev,~\IEEEmembership{Member,~IEEE}% <-this % stops a space
\thanks{
Manuscript received August 10, 2021; revised October 04, 2021; revised November 13, 2021; and accepted December 03, 2021.
}

\thanks{
B.\ McNicholl, A.\ Campbell and S.\ Dev are with University College Dublin, Ireland (e-mail: barry.mcnicholl@ucdconnect.ie, abey.campbell@ucd.ie, soumyabrata.dev@ucd.ie). Y.\ H.\ Lee is with the School of Electrical and Electronic Engineering, Nanyang Technological University, Singapore (e-mail: EYHLee@ntu.edu.sg).
}% <-this % stops a space
\thanks{
Send correspondence to S.\ Dev, E-mail: soumyabrata.dev@ucd.ie.}% <-this % stops a space
\thanks{
The ADAPT Centre for Digital Content Technology is funded under the SFI Research Centres Programme (Grant 13/RC/2106\_P2) and is co-funded under the European Regional Development Fund.}
}

\markboth{IEEE Geoscience and Remote Sensing Letters,~Vol.~XX, No.~XX, XX~2021}%
{Shell \MakeLowercase{\textit{et al.}}: Bare Demo of IEEEtran.cls for IEEE Journals}



\maketitle

\begin{abstract}
The reliability of ERA5 satellite-based air temperature data is under investigation in this paper. To evaluate this, the ERA5 data will be compared with land-based data obtained from weather stations on the Global Historical Climatology Network. Two climate regions are taken into consideration, temperate and tropical. Five years worth of data is collected and compared through box plots, regression models and statistical metrics. The results show that the satellite temperature performs better in the temperate region than the tropical region. This suggests that the time of year and climate region have an impact on the accuracy of the satellite data as milder temperatures produce better approximations.
\end{abstract}

% Note that keywords are not normally used for peer review papers.

\IEEEpeerreviewmaketitle

\section{Introduction}
\label{sec:intro}
\IEEEPARstart{T}{he} analysis of climate is an essential process in the modern world. By documenting weather conditions over time, predictions of future weather patterns can be made around the world~\cite{manandhar2019data}. This impacts various fields including agriculture, tourism, and renewable energy \cite{becken2010importance}. Due to the dependence of society on climate analysis, reliable weather measurements are needed. Air temperature is one of the weather variables that is extensively recorded and estimated~\cite{wang2021day}. It can be measured on the surface of the Earth or from satellites.

ERA5 is an atmospheric reanalysis of the world's climate using satellite measurements and is provided by the European Centre for Medium-Range Weather Forecasts (ECMWF)\footnote{\label{note1}ERA5 data is available at \url{https://cds.climate.copernicus.eu/cdsapp\#!/dataset/reanalysis-era5-single-levels?tab=overview}}. Reanalysis incorporates physical laws to combine observations and model data into a comprehensive data set. To analyze land surface variables such as air temperature, ERA5 uses a land data assimilation system which is weakly coupled with a 4D variational data assimilation  \cite{hersbach2020era5}. ERA5 has replaced ERA Interim reanalysis and offers several improvements such as better performance over land \cite{wang2019comparison}. It also provides high spatial and temporal resolution and offers data which spans back several decades. All of these factors suggest that ERA5 is a trust worthy source of temperature data. However, there is little proof of this to date. For this reason, this paper will focus on determining the reliability of air temperature data provided by ERA5.

Temperature data provided by the Global Historical Climatology Network (GHCN) will act as a comparison to the ERA5 data to determine its reliability. The GHCN is a database of climate summaries obtained from land stations across the globe\footnote{\label{note2}Data from the GHCN is available at \url{https://www.ncdc.noaa.gov/data-access/land-based-station-data/land-based-datasets/global-historical-climatology-network-ghcn}}. The GHCN is provided by the National Oceanic and Atmospheric Administration (NOAA). Data from multiple sources is integrated into the GHCN after it has been screened and each station has been classified. Variables such as air temperature are obtained from over 100,000 stations across the world where some data originates as far back as 1763 \cite{jaffres2019ghcn}. Temperature measurements on the GHCN are generally considered to very meticulous based on quality control tests which have been carried out \cite{peterson1998global}. The proven reliability of GHCN is why it will be used as a source of control data for the analysis of ERA5.

Comparisons between ERA5 and GHCN measurements have been made in the past for precipitation in the Northeastern United States. This study found that distance from the coast and elevation affected the precipitation relationship between the two data sets \cite{crossett2020evaluation}. These results suggest that differences between air temperature measurements will also be found between the data sets. These differences will be investigated for two different climate types for multiple years worth of data (2015-2019) to provide a comprehensive picture of the accuracy of ERA5. Dublin will be used as a temperate region and Singapore as a tropical region. The main contributions of this paper can be summarized as follows:

\begin{itemize}
    \item Comparison of satellite data estimations to land data;
    \item Determination of possible causes of estimation errors for air temperature;
    \item Evaluation of the reliability of ERA5 for estimating air temperature.
\end{itemize}

The rest of the paper\footnote{\label{note3}In the spirit of reproducible research, the code used to collect and analyze data sets is available at \url{https://github.com/Barry-McNicholl/sat-land-temp}} is structured as follows. Section II discusses the temperature data sets, how they are obtained and compared. Section III quantifies the difference in temperature between the data sets. Section IV evaluates the land temperature estimates from the satellite temperatures. Section V sums up the conclusions and future work.

\section{Data \& Methods}
\label{sec:data-&-methods}
\subsection{Global Historical Climatology Network Data}
\label{subsec:ghcn-data}
The GHCN land-based data can be sourced from the NOAA website. Specifically, the data set of daily climate summaries was used in this work. As Dublin had been selected for the temperate region, a station in Dublin had to be selected to extract land data from. The station chosen for this region is the only available station in Dublin and is situated in Phoenix Park. Similarly, only one station was available for the tropical climate in Singapore. This station is located in the Singapore Changi International Airport.

The data is downloaded using an API and stored in JSON format~\cite{pathan2021analyzing}. Temperature values were requested for every day between the years 2015 and 2019. For land air temperature values, the variable used was TAVG which is the average temperature over the course of a day measured from 00:00 on one day to 00:00 on the next day. Each data point has an associated timestamp which corresponds to the day the measurement was taken. Using this timestamp, the data can be separated into lists where data from the same months are kept together and months from the same year. This was done so that the data could be analyzed more easily.

\subsection{ERA5 Data}
\label{subsec:era5-data}
Satellite-based data was sourced from the Climate Data Store (CDS) website which was implemented by the ECMWF. The data set of interest consists of hourly ERA5 data on single levels and dates back to 1979. As previously mentioned the primary use for the satellite data is to analyze its reliability and compare it to the land data. However, it will also be used to create a visual representation of the ERA5 temperature distribution for both regions. An initial judgement on ERA5's reliability can be made here depending on whether the temperature is spread evenly or not.

The data is obtained through the use of an API and is stored in netCDF format. When downloading satellite data, the geographical area must be specified in latitude and longitude coordinates. Four coordinates are selected which are the most northern, southern, eastern and western edges of the area. The number of downloaded data points is dependent on the chosen coordinates. If the coordinates are far apart, there will be a greater number of points between the coordinates. In an attempt to minimize the number of data points, the coordinates for each region were chosen so that only 9 data points were downloaded. This number of points is sufficient for viewing the temperature distribution of the regions whilst also reducing the time taken to download the data. The data point at the centre of each region is as close as possible to the land stations, so this point will be used for comparison with the land data.

The variable used for satellite air temperature here is 2m temperature which is the temperature of the air measured at 2 metres above the ground at a given point. Average daily data was not available here so temperatures were downloaded for every hour of each day from 2015-2019 and the average of each day is calculated. Once the averages are calculated, timestamps are used to store the data like before.

\subsection{Comparison of Data Sets}
\label{subsec:data-set-comp}

The most straight forward method of comparing the temperature values from the data sets is to simply subtract the land values for each day from the satellite values. The temperature difference between the land and satellite data gives an indication of how closely related the data sets are to each other. An effective way of illustrating these temperature differences is to group them by month and create box plots. This will highlight the variance of the data, whether there are many outliers and so on. Only days where both land measurements and satellite measurements were available were considered when calculating the temperature differences.

A linear regression is used to subjectively evaluate the data sets with land temperature on one axis and satellite temperature on the other. For this, all of the satellite data is combined into one data set. The same applies for the land data. Various metrics, will be used to compare the data sets. These metrics are R\textsuperscript{2} coefficient, Kling-Gupta Efficiency (KGE), Mean Bias Error (MBE), Root Mean Square Error (RMSE) and Dynamic Time Warping (DTW) distance:

\begin{equation}
\label{eqn:equation1}
R^2=1-\frac{\sum_{i=0}^{n}(\hat{y}_i-y_i)^2}{\sum_{i=0}^{n}(y_i-\bar{y})^2}
\end{equation}
\begin{equation}
\label{eqn:equation2}
KGE=1-\sqrt{(r-1)^2+(\frac{\sigma_{\hat{y}}}{\sigma_{y}}-1)^2+(\frac{\mu_{\hat{y}}}{\mu_{y}}-1)^2}
\end{equation}
\begin{equation}
\label{eqn:equation3}
MBE=\frac{1}{n}\sum_{i=0}^{n}(\hat{y}_i-y_i)
\end{equation}
\begin{equation}
\label{eqn:equation4}
RMSE=\sqrt{\frac{1}{n}\sum_{i=0}^{n}(\hat{y}_i-y_i)^2}
\end{equation}
\begin{equation}
\label{eqn:equation5}
%\begin{aligned}
\begin{array}{c}
Distance=d(n,m) \\
d(i,j)=||y_i-\hat{y}_j||+\min
\begin{cases}
d(i,j-1) \\
d(i-1,j) \\
d(i-1,j-1) \\
\end{cases} \\
d(0,0)=0, d(i,0)=d(0,j)=\infty \\
(i=1,...,n; j=1,...,m) \\
\end{array}
%\end{aligned}
\end{equation}

In the equations above, n is the number of data samples, y is the measured value from the land data, \^{y} is the predicted value from the satellite data, \textmu\textsubscript{y} is the mean of y, \textmu\textsubscript{\^{y}} is the mean of \^{y}, \textsigma\textsubscript{y} is the standard deviation of y, \textsigma\textsubscript{\^{y}} is the standard deviation of \^{y} and r is the linear correlation between y and \^{y}.

To provide an objective evaluation, supervised machine learning regression models can be employed to estimate the land-based temperature from the satellite-based temperature. Various models will be used to cross validate the data where the land and satellite values are both separated by year. One of the years acts as a test data set whereas the other four are used as a training data set. This approach produces three metrics, the R\textsuperscript{2} coefficient, RMSE and time taken to use the model.

\section{Variation in Air Temperature}
\begin{figure}[htb]
\begin{center}
\includegraphics[width=0.35\textwidth]{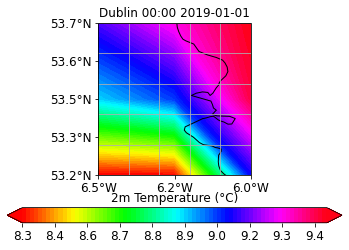}\\
\includegraphics[width=0.35\textwidth]{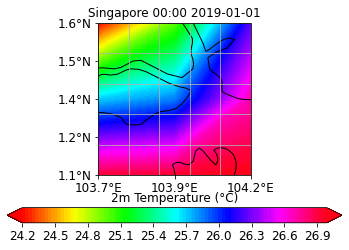}
\caption{We visualize the satellite temperature distribution for Dublin and Singapore on 00:00 01/01/19.
\label{fig:dublin-singapore-temp-dist-19}}
\end{center}
\end{figure}

The distribution of temperature based on satellite measurements, on a given time and date, can be seen for Dublin and Singapore in \autoref{fig:dublin-singapore-temp-dist-19}. From these images, it is clear that in the surrounding areas of Dublin and Singapore, the temperature values are spread evenly. The even spread of temperatures shows that the temperature data produced by ERA5 is potentially reliable, as opposed to a data set with large fluctuations between data points. The high spatial resolution of ERA5 means that temperatures can potentially be mapped out accurately over relatively small geographical areas. This would make it a valuable data set, which has been noted in its use to support hydrological studies \cite{munoz2021era5}.

\begin{figure}[htb]
\begin{center}
\includegraphics[height=0.25\textwidth]{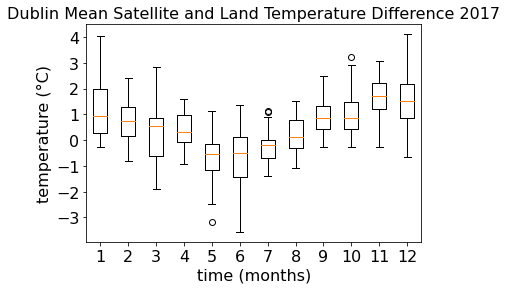}\\
\includegraphics[height=0.25\textwidth]{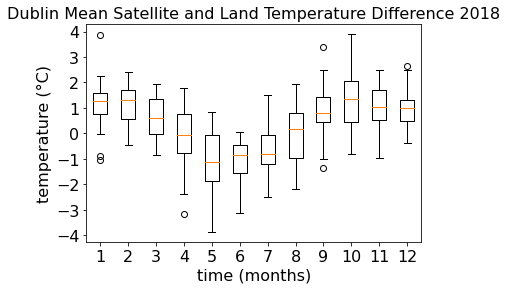}\\
\includegraphics[height=0.25\textwidth]{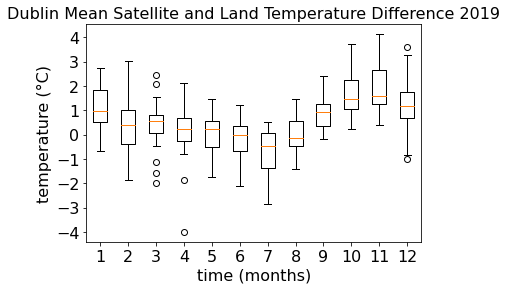}
\caption{We demonstrate the temperature difference between satellite and land temperature in Dublin for 2017-2019.
\label{fig:dublin-temp-diff-2016-2019}}
\end{center}
\end{figure}

The box plots for Dublin from 2017-2019 are shown in \autoref{fig:dublin-temp-diff-2016-2019}. It is clear that for each year, the temperature difference varies in a cyclic pattern. This can be observed when following the median of each box which is marked by an orange line. The difference is at its lowest point during the summer months and it peaks during the winter months. This means that the satellite temperature is underestimated in the summer and overestimated in the winter. The most accurate predictions occur in the months in between these two seasons, where the median temperature difference is closer to zero.

\begin{figure}[htb]
\begin{center}
\includegraphics[height=0.25\textwidth]{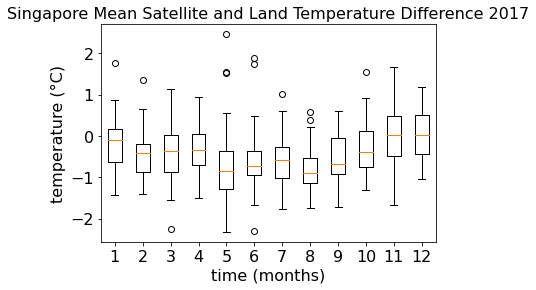}\\
\includegraphics[height=0.25\textwidth]{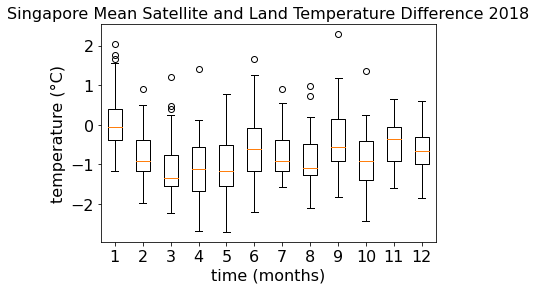}\\
\includegraphics[height=0.25\textwidth]{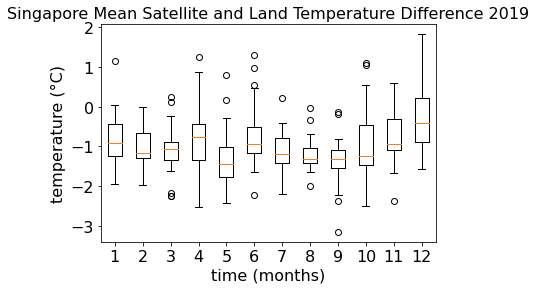}
\caption{We demonstrate the temperature difference between satellite and land temperature in Singapore for 2017-2019.
\label{fig:singapore-temp-diff-2016-2019}}
\end{center}
\end{figure}

\autoref{fig:singapore-temp-diff-2016-2019} contains the box plots for Singapore from 2017-2019. Unlike Dublin, there is no clear cyclic pattern followed by the temperature difference each year. Here, the satellite temperature appears to be underestimated across all the months but the amount it is underestimated by is arbitrary. The mean air temperature for each month in Singapore is roughly even so this is likely the cause for the arbitrary nature of the data.

In general, it looks like there is roughly an equal amount of variance in the boxes between Dublin and Singapore with both regions having similar interquartile ranges and numbers of outliers, which are marked by circles. From the box plots for both regions, it seems that the temperature difference is influenced by how high or low the air temperature is when measurements are being taken. Underestimation occurs when the air temperature is relatively high which would be common in Dublin during the summer and Singapore throughout the year. On the other hand, overestimation is caused by relatively low air temperatures which would occur in Dublin during the winter. The temperatures differences appear to be closest to 0 when the mean temperature for the month is roughly 10 degrees celsius, which occurs in Dublin around April and September. ERA5 is known to have a bias when compared to land based measurements \cite{liu2021well}. The results in this work show that there is a positive bias in the winter and a negative bias in the summer after post processing has been carried out.

% All years
% \begin{figure*}[htb]
% \begin{center}
% \includegraphics[height=0.2\textwidth]{Dublin_Boxplot_2015.png}
% \includegraphics[height=0.2\textwidth]{Dublin_Boxplot_2016.png}
% \includegraphics[height=0.2\textwidth]{Dublin_Boxplot_2017.png}
% \includegraphics[height=0.2\textwidth]{Dublin_Boxplot_2018.png}
% \includegraphics[height=0.2\textwidth]{Dublin_Boxplot_2019.png}
% \caption{Dublin Temperature Difference 2015-2019.
% \label{fig:dublin-temp-diff-2015-2019}}
% \end{center}
% \end{figure*}

% All years
% \begin{figure*}[htb]
% \begin{center}
% \includegraphics[height=0.2\textwidth]{Singapore_Boxplot_2015.png}
% \includegraphics[height=0.2\textwidth]{Singapore_Boxplot_2016.png}
% \includegraphics[height=0.2\textwidth]{Singapore_Boxplot_2017.png}
% \includegraphics[height=0.2\textwidth]{Singapore_Boxplot_2018.png}
% \includegraphics[height=0.2\textwidth]{Singapore_Boxplot_2019.png}
% \caption{Singapore Temperature Difference 2015-2019.
% \label{fig:singapore-temp-diff-2015-2019}}
% \end{center}
% \end{figure*}

\section{Estimating Land Temperature from Satellite Temperature Measurements}
\subsection{Subjective Evaluation}

\begin{figure}[htb]
\begin{center}
\includegraphics[width=0.4\textwidth]{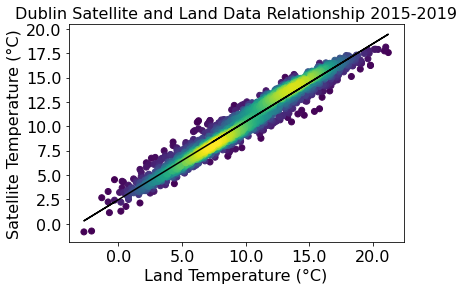}\\
\includegraphics[width=0.4\textwidth]{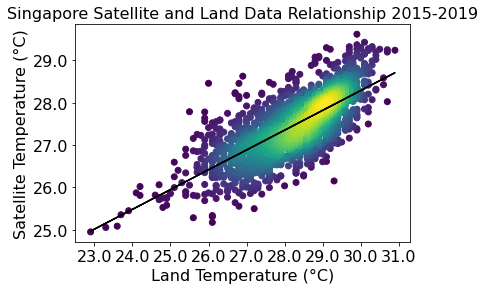}
%\caption{Dublin \& Singapore Linear Regression 2015-2019.
\caption{We compute the linear regression between satellite temperature and land temperature of Dublin \& Singapore across the years 2015-2019.
\label{fig:dublin-singapore-linear-reg-2015-2019}}
\end{center}
\end{figure}

\autoref{fig:dublin-singapore-linear-reg-2015-2019} shows the linear regression for Dublin from 2015-2019. The land and satellite temperatures are highly correlated in this instance. This conclusion can be drawn from the fact that in \autoref{tab:table1}, the R\textsuperscript{2} coefficient and KGE are relatively high, and that the MBE, RMSE and DTW distance are low. Hence, the satellite temperature provides a good estimate of the land temperature in this case.

\begin{table}[htb]
\begin{center}
\caption{We calculate the various metrics for Dublin \& Singapore linear regression across the years 2015-2019.}
\label{tab:table1}
\begin{tabular}{l|r|r}
\toprule
                              & \multicolumn{1}{c|}{Dublin} & \multicolumn{1}{c}{Singapore} \\
\midrule
Metric                        & \multicolumn{1}{l|}{Value}  & \multicolumn{1}{l}{Value}     \\
\midrule
%Best Fit Line Equation        & y = 0.8014x + 2.4586       & y = 0.4674x + 14.2641         \\
Dynamic Time Warping Distance & 1845.24                  & 1429.39                     \\
Kling-Gupta Efficiency        & 0.72                     & 0.41                        \\
Mean Bias Error               & 0.49                     & -0.75                       \\
R-Squared                     & 0.91                     & 0.11                        \\
Root Mean Square Error        & 1.30                     & 1.10                        \\
\bottomrule
\end{tabular}
\end{center}
\end{table}

The Singapore linear regression for the same period is also visible in \autoref{fig:dublin-singapore-linear-reg-2015-2019}. Here, the satellite temperature does not provide as good of an estimate of the land temperature. This is clear from \autoref{tab:table1} where there is a lower R\textsuperscript{2} coefficient and KGE. The MBE is a relatively large negative number which indicates that satellite values are underestimated compared to the land values. Like the temperate region, the RMSE and DTW distance are relatively low. However, this is likely due to the fact that various outliers are present in the data, which are skewing the results. We also compute the best fit line equation for Dublin and Singapore as:

\begin{equation}
\begin{aligned}
y = 0.80x + 2.46 \mbox{ (For Dublin)}\\
y = 0.47x + 14.26 \mbox{ (For Singapore)},
\end{aligned}
\end{equation}

where for each region, $x$ and $y$ represents the land and satellite temperatures respectively.

\begin{table*}[htb]
\begin{center}
\caption{Estimation of land temperature from satellite temperature measurements using 2015 as test data, using several machine learning based methods.}
\label{tab:table2}
\begin{tabular}{l|rrr|rrr}
\toprule
                              & \multicolumn{3}{c|}{Dublin}       & \multicolumn{3}{c}{Singapore}     \\
\midrule
Model                         & R-Squared & RMSE & Time Taken & R-Squared & RMSE & Time Taken \\
\midrule
AdaBoostRegressor             & 0.94    & 0.84  & 0.05     & 0.49     & 0.54  & 0.03     \\
BaggingRegressor              & 0.94    & 0.80  & 0.03     & 0.51     & 0.53  & 0.02     \\
BayesianRidge                 & 0.95    & 0.78  & 0.01     & 0.50     & 0.54  & 0.01     \\
DecisionTreeRegressor         & 0.94    & 0.80  & 0.01     & 0.50     & 0.53  & 0.01     \\
DummyRegressor                & -0.01   & 3.38  & 0.01     & -0.08    & 0.79  & 0.01     \\
ElasticNet                    & 0.77    & 1.61  & 0.01     & -0.03    & 0.77  & 0.01     \\
ElasticNetCV                  & 0.95    & 0.78  & 0.08     & 0.50     & 0.54  & 0.08     \\
ExtraTreeRegressor            & 0.94    & 0.82  & 0.01     & 0.51     & 0.53  & 0.01     \\
ExtraTreesRegressor           & 0.94    & 0.80  & 0.20     & 0.51     & 0.53  & 0.14     \\
GaussianProcessRegressor      & 0.95    & 0.76  & 0.30     & 0.51     & 0.53  & 0.32     \\
GradientBoostingRegressor     & 0.95    & 0.77  & 0.08     & 0.51     & 0.53  & 0.07     \\
HistGradientBoostingRegressor & 0.95    & 0.77  & 0.89     & 0.51     & 0.53  & 0.93     \\
HuberRegressor                & 0.95    & 0.78  & 0.02     & 0.50     & 0.54  & 0.03     \\
\bottomrule
\end{tabular}
\end{center}
\vspace{-0.5cm}
\end{table*}

The higher correlation in Dublin, as opposed to Singapore, seems to be as a result of milder temperatures in Dublin. This matches the observations of underestimation and overestimation that were made whilst analyzing the box plots.

\subsection{Objective Evaluation}

The results for Dublin and Singapore with 2015 as the test data set can be seen in \autoref{tab:table2}. For Dublin, in the majority of models, the R\textsuperscript{2} coefficient and RMSE are relatively high and low respectively. These values are also quite similar to the values obtained in the subjective evaluation. Therefore, like the subjective evaluation, the land temperature can still be accurately estimated from the satellite temperature for the temperate region.

For Singapore, most models offer a moderate R\textsuperscript{2} coefficient along with a low RMSE. The low RMSE can again be explained by the presence of outliers. This is also in line with the subjective evaluation, which indicates that the land temperature can not be predicted as accurately from the satellite temperature for the tropical region when compared to the temperate region.

The reason for the higher correlation in Dublin when compared to Singapore can again be explained by the milder temperatures in Dublin. This result of higher and lower correlations due to overestimation and underestimation of the satellite data is as expected from the observations made earlier. These findings are supported by a study of the bias in ERA5 air temperature over the Iberian peninsula \cite{johannsen2019cold}. The results in this work indicate that daytime temperatures in the summer are underestimated and temperatures at night are overestimated.

\section{Conclusions}
\label{sec:conclusion}
There is no doubt that ERA5 data is useful for providing estimates of weather conditions in areas where no land measurements are available \cite{hoffmann2019era}. The question marks surrounding ERA5 comes from its reliability when compared to land measurements. The temperature difference found between the satellite and land data in this paper proved to be relatively large in some cases. From the findings, the satellite data provided a better estimate of the land data for the temperate region than the tropical region.

The scope of this study could be expanded upon in the future to include data from additional regions with different climates. This would offer more insight into the reliability of ERA5 temperature values and could strengthen the findings already presented in this work. To get an overall picture of the value of ERA5, multiple weather variables should be analyzed and compared with land values. It is possible that some variables might be more highly correlated than others. However, in terms of temperature, ERA5 data is most reliable when measuring mild temperatures (close to 10 degrees celsius) which are most common in temperate regions at certain times of the year. Land measurements are more suitable in other scenarios where temperature is high or low. This includes hot or cold climates and times of the year when the air temperature may be low or high.

\vspace{-0.6cm}

%\bibliographystyle{IEEEbib}
%\bibliographystyle{IEEEtran}

\end{document}